\documentclass[9pt,conference]{IEEEtran}
\IEEEoverridecommandlockouts
\usepackage{cite}
\usepackage{amsmath,amssymb,amsfonts}
\usepackage{algorithmic}
\usepackage{graphicx}
\usepackage{textcomp}
\usepackage{tabularx}
\usepackage{xcolor}
\usepackage{multirow}
\usepackage{graphicx}
\usepackage{amsmath}
\usepackage{xcolor}
\usepackage{afterpage}
\usepackage{svg}
\usepackage{hyperref}
\usepackage{balance}

\def\BibTeX{{\rm B\kern-.05em{\sc i\kern-.025em b}\kern-.08em
    T\kern-.1667em\lower.7ex\hbox{E}\kern-.125emX}}
\begin{document}

\title{Mel-Refine: A Plug-and-Play Approach to Refine Mel-Spectrogram in Audio Generation 
}
\author{\IEEEauthorblockN{Hongming Guo\textsuperscript{1}, Ruibo Fu\textsuperscript{2}, Yizhong Geng\textsuperscript{1}, Shuai Liu\textsuperscript{1}, Shuchen Shi\textsuperscript{2}, Tao Wang\textsuperscript{2}, Chunyu Qiang\textsuperscript{2}, Chenxing Li\textsuperscript{4}\\ Ya Li\textsuperscript{1}, Zhengqi Wen\textsuperscript{3}, Yukun Liu\textsuperscript{2}, Xuefei Liu\textsuperscript{2}}
 \IEEEauthorblockA{
       \textsuperscript{1}Beijing University of Posts and Telecommunications, Beijing, China\\
       \textsuperscript{2}Institute of Automation, Chinese Academy of Sciences, Beijing, China\\
        \textsuperscript{3}Beijing National Research Center for Information Science and Technology, Tsinghua University, Beijing, China\\
        \textsuperscript{4}Tencent AI Lab, Shenzhen, China\\
        Emails: \{ghm0221@bupt.edu.cn, ruibo.fu@nlpr.ia.ac.cn\}
    }
}

\maketitle

\begin{abstract}
Text-to-audio (TTA) model is capable of generating diverse audio from textual prompts. However, most mainstream TTA models, which predominantly rely on Mel-spectrograms, still face challenges in producing audio with rich content. The intricate details and texture required in Mel-spectrograms for such audio often surpass the models' capacity, leading to outputs that are blurred or lack coherence. In this paper, we begin by investigating the critical role of U-Net in Mel-spectrogram generation. Our analysis shows that in U-Net structure, high-frequency components in skip-connections and the backbone influence texture and detail, while low-frequency components in the backbone are critical for the diffusion denoising process. We further propose ``Mel-Refine'', a plug-and-play approach that enhances Mel-spectrogram texture and detail by adjusting different component weights during inference. Our method requires no additional training or fine-tuning and is fully compatible with any diffusion-based TTA architecture. Experimental results show that our approach boosts performance metrics of the latest TTA model Tango2 by 25\%, demonstrating its effectiveness.
Project Page: \href{https://github.com/hongming21/Mel-Refine}{https://github.com/hongming21/Mel-Refine}
\end{abstract}


\section{Introduction}
Recently TTA\cite{2000hmm, echoaudio, auffusion, makean, huang2023make2, diffava} model has gained significant attention, which can greatly assist in the production of multimedia content. Mainstream TTA model can be broadly categorized into two main types. The first type converts audio into discrete tokens\cite{borsos2023audiolm,liu2024audioldm,rubenstein2023audiopalm,deshmukh2023pengi,yang2024uniaudio}, followed by using a transformer model, with AudioGen\cite{kreuk2023audiogen} and MusicGen\cite{musicgen} being representative examples. 

The second type transforms audio into a Mel-spectrogram and generates audio by decoding the generated Mel-spectrogram. Since the dimensionality of Mel-spectrograms is similar to images, these methods often draw on models and techniques from the image generation field, producing Mel-spectrograms in a manner similar to image synthesis. Representative works include DiffSound\cite{yang2023diffsound}, which employs vector quantization variational autoencoder (VQ-VAE)\cite{rombach2022high} for autoregressive generation, LAFMA\cite{guan2024lafma}, which utilizes ODE-based diffusion processes, and the Audioldm\cite{audioldm1,audioldm2} and Tango\cite{tango,mustango,majumder2024tango} family of models, which are based on architectures similar to latent diffusion model (LDM) \cite{rombach2022high}. Among these, LDM-based models have achieved the widest application and best performance, largely due to the diffusion model's ability to model high-dimensional continuous variables. Trained on large-scale text-audio pairs, these models possess the ability to generate music, vocals, environmental sounds, and complex audio scenes in alignment with the provided textual prompts.

\begin{figure}[t]
    \centering
    \includegraphics[width=\columnwidth]{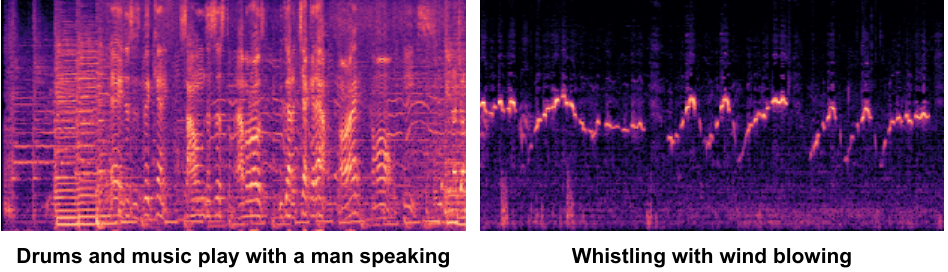} 
    \caption{Comparison of Mel-spectrograms between simple and complex audio scenarios. The Mel-spectrogram corresponding to rich content audio scenarios contains more details and textures, which poses a challenge to the model's ability to capture and represent them.}
    \label{fig:1}
\end{figure}

While these models have significantly enhanced the alignment between generated audio and the provided text, audio quality still requires further improvement. When the text prompt becomes more intricate, incorporating multiple sound events or themes, the generated audio tends to lose clarity and coherence. This suggests that these models still struggle to model complex acoustic scenes. This is primarily because complex audio scenes correspond to detailed and intricate Mel-spectrograms, where various sound components must be accurately represented in terms of their frequency and decibel changes over time, as shown in Fig. \ref{fig:1}. Inaccurate representation of these components leads to muffling and distortion in the generated audio. Although current TTA models can model the general structure of Mel-spectrograms, they still struggle with precise modeling of finer details, leading to reduced performance in generating rich content audio.

Recently, the FreeU\cite{si2024freeu} method reveal the uncover potential of U-Net\cite{unet} architectures for denoising within diffusion models. Their research shows that enhancing the backbone can improve the model's denoising ability, but it may lead to overly smooth generated images. Reducing the low-frequency components of the skip connection can effectively mitigate the excessive smoothness. Building on this, we explored the contributions of different components within the U-Net to Mel-spectrogram generation. We find that the high-frequency components of both the skip connections and backbone in the U-Net structure impact the modeling of fine details and edges in the generated Mel-spectrogram. Enhancing the high-frequency components of the skip connections or suppressing those of the backbone can both improve the model's ability to capture the intricate structure and edges of the Mel-spectrogram. Conversely, enhancing low-frequency components of the backbone can lead to a considerable degradation in the model's denoising performance. Based on these findings, we propose the ``Mel-Refine'' method, which requires no additional training or fine-tune. By simply adjusting the weight of different components of the U-Net structure during inference. This method significantly improves the model's ability to capture the complexity of the generated Mel-spectrogram, improving audio quality in both objective metrics and subjective perception.

Overall, this paper makes the following three contributions:

1. We validate the distinct roles of U-Net components in generating Mel-spectrograms, identifying that high-frequency components in both skip-connection and backbone impact the detail and texture of generated Mel-spectrograms, while the low-frequency components in backbone are mainly related to the model's denoising ability.

2. We propose ``Mel-Refine'', a plug-and-play approach that enhances generation quality during inference by adjusting the influence of different components through the tuning of a few parameters.

3. The proposed ``Mel-Refine'' framework is versatile and seamlessly integrates with existing LDM-based audio generation models. We demonstrate its effectiveness through both objective metric improvements and enhanced subjective perception.

\afterpage{
\begin{figure*}[t]
    \centering
    \includegraphics[width=0.8\textwidth]{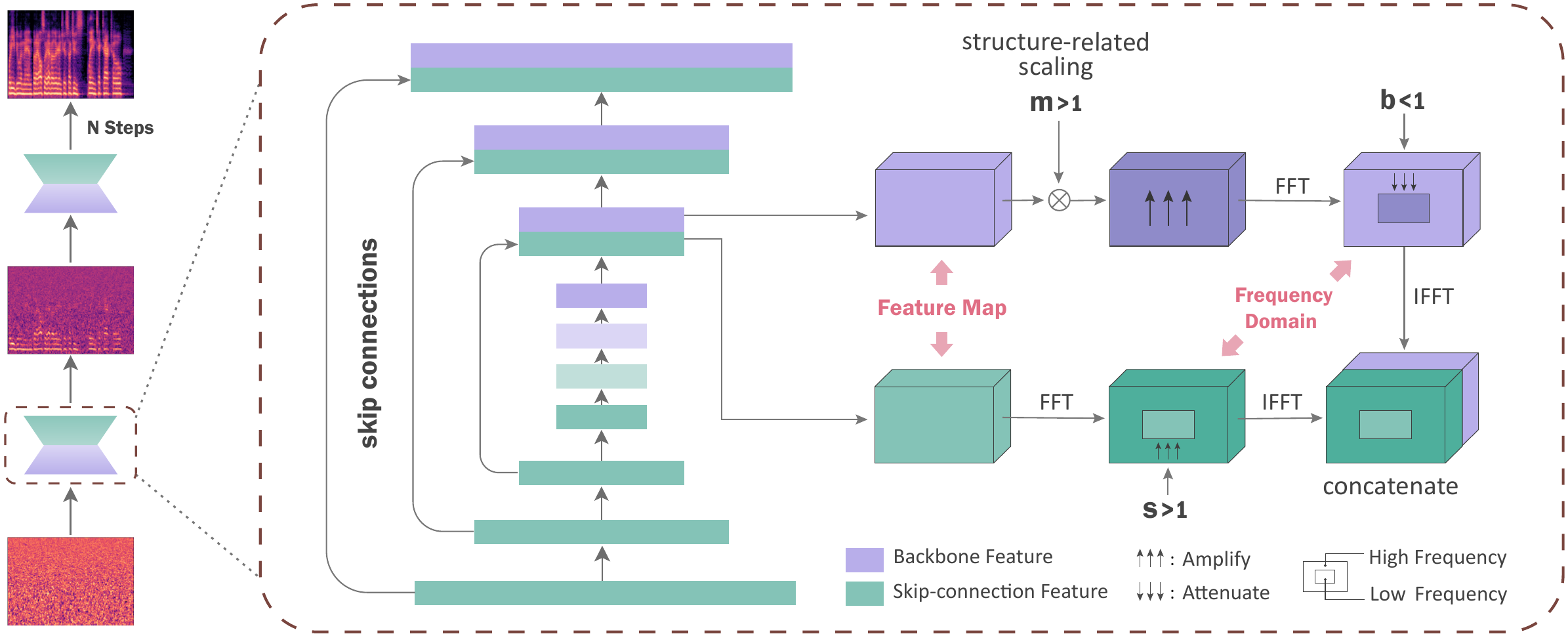} 
    \caption{The whole framework of Mel-Refine}
    \label{fig:3}
\end{figure*}}

\section{Methodology}
In this section, we first explore the impact of different U-Net components on Mel-spectrogram generation and then propose our method, Mel-Refine, to enhance the generated audio quality.

\subsection{Impact of U-Net  
Different Components on Generation}

\textbf{Preliminary:} U-Net is a encoder-decoder model that stands out for its use of skip-connections. These connections link features from corresponding encoder layers directly to the decoder layers by concatenating them. This design helps preserve spatial information and improves the model's ability to reconstruct fine details during the decoding process.

\textbf{Motivation:} Inspired by FreeU, we investigate the impact of different U-Net components on Mel-spectrogram generation. FreeU suggests that enhancing the U-Net backbone improves denoising performance but may cause over-smoothing in the generated images. To mitigate this, suppressing low-frequency components in the skip-connections helps reduce over-smoothing and enhances texture. Based on this, we divide the U-Net into four key segments: high-frequency and low-frequency elements in both the skip connections and the backbone. By adjusting these components, we analyze their effects on the final generated Mel-spectrogram.

\begin{figure}[t]
    \centering
    \includegraphics[width=\columnwidth]{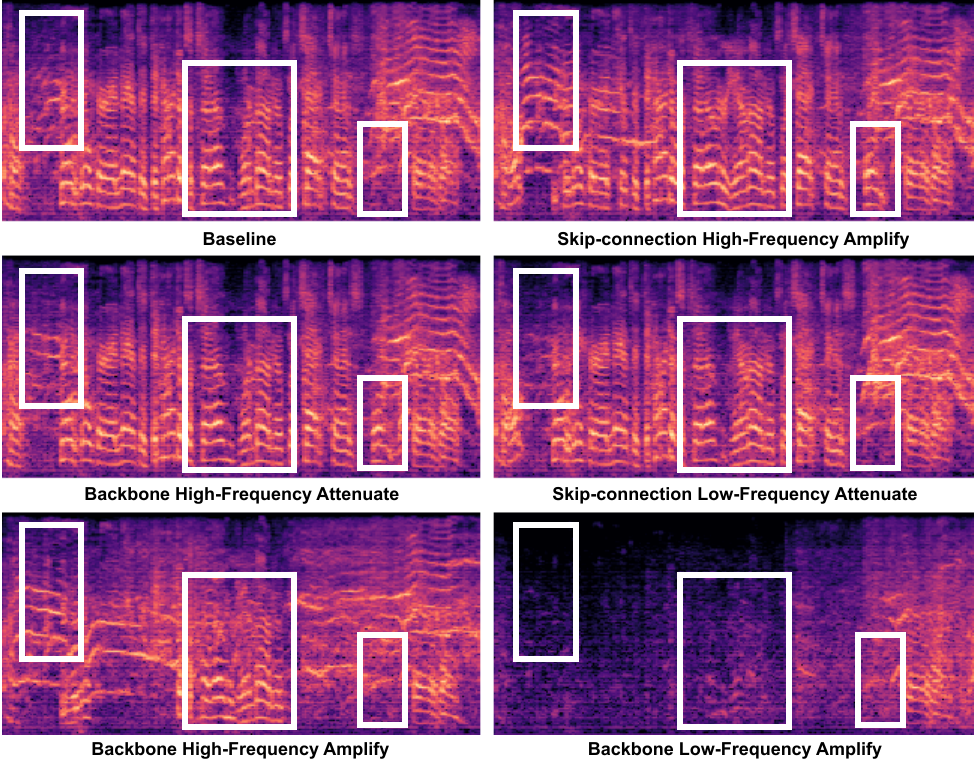} 
    \caption{This figure illustrates the impact of altering weights at different frequencies on the generated Mel spectrogram. The baseline refers to the output from the Tango2 model without any modifications. The text prompt used is ``A man talking followed by screaming children, followed by more high-pitched conversation.'' The figure clearly demonstrates that enhancing the high-frequency components of the skip connections introduces more textural details. In contrast, amplifying the low-frequency components of the backbone results in significant model degradation. Suppressing the high-frequency components of the backbone yields a slight improvement.}
    \label{fig:fig2}
\end{figure}

\textbf{Experiments:} We employ Fourier transformation to extract the high-frequency and low-frequency components, by scaling these components with coefficients greater or less than 1, we amplify or attenuate their contribution. We conducted experiments using the latest Tango2\cite{majumder2024tango} model, adjusting the influence of different components to observe the variations in the generated Mel-spectrograms and calculate objective metrics against the reference audio. The results are presented in Table. \ref{tab:small_exp} and Fig.\ref{fig:fig2}. Results indicate that \textbf{amplifying the high-frequency components in the skip-connection} significantly improves the model's performance, while \textbf{attenuating the low-frequency components in the backbone} has a substantial negative impact on the generation quality. Additionally, attenuating both low-frequency components in the skip-connection and high-frequency components in the backbone also leads to a slight improvement in the results, but the skip-connection low-frequency components modeification shows more noticeable improvements on the Mel spectrogram.

\textbf{Explanation:} We hypothesize this result arises from the differing roles of the backbone and skip-connection in the diffusion denoising process. The backbone primarily handles noise predict, and thus, the amplification of low-frequency components may lead to excessive denoising, removing high-frequency details and textures along with the noise, which degrades the model's output. In contrast, the skip-connection primarily transmits structural information, and enhancing high-frequency components helps preserve details and textures, allowing them to be better retained during the regeneration process.

\begin{table}[htbp]
\caption{Objective Result of Different Components Impact}
\begin{center}
\resizebox{\columnwidth}{!}{ 
\begin{tabular}{|c|c|c|c|c|}
\hline
\multirow{2}{*}{\textbf{Operation}} & \multicolumn{2}{|c|}{\textbf{Skip-connection}} & \multicolumn{2}{|c|}{\textbf{Backbone}} \\
\cline{2-5} 
 & \textbf{\textit{High-frequency}} & \textbf{\textit{Low-frequency}} & \textbf{\textit{High-frequency}} & \textbf{\textit{Low-frequency}} \\
\hline
\multirow{1}{*}{Amplify} & \textbf{34.31/2.11} &36.09/2.44 &31.22/2.70 & 37.28/2.56 \\
\hline
\multirow{1}{*}{Attenuate} & 37.53/2.54 & 35.71/2.38 &36.82/2.32 & \textbf{36.57/9.58} \\
\hline
\multicolumn{5}{l}{All values represent FD/FAD, where lower values are better. Baseline is \textbf{37.48/2.48}}
\end{tabular}
} 
\label{tab:small_exp}
\end{center}
\end{table}

\subsection{Mel-Refine}
Based on previous finding, We propose our method named ``Mel-Refine'', to enhance the audio quality generated by the TTA model. This approach improves the textural details and overall quality of the generated Mel-spectrogram by adjusting the high-frequency weights of the skip connections and the backbone during inference, without requiring additional training or fine-tuning. The framework is illustrated in Fig. \ref{fig:3}.

In U-Net decoder, let $x$ represent the backbone feature map and $h$ the skip-connection feature map. Recognizing that features become sparser in later decoder stages, we apply the following operations only in the first two decoder blocks.

\textbf{For the skip-connection features}, we first apply a Fourier transform to the feature map, enhance its high-frequency components by scaling them with a parameter \( s > 1 \), and then return to the spatial domain using the inverse Fourier transform. Given that the skip-connection feature map has a shape of \( (B, C, H, W) \), the process can be described as follows:

\begin{equation}
    \mathcal{F}(h) = \text{FFT}(h),
\end{equation}

\begin{equation}
    \mathcal{F}'(h) = \mathcal{F}(h) \odot \beta,
\end{equation}

\begin{equation}
    h' = \text{IFFT}(\mathcal{F}'(h)).
\end{equation}

Here, \(\odot\) denotes element-wise multiplication, and \(\beta(x, y)\) is a scaling matrix defined as:

\begin{equation}
\beta(x, y) = 
\begin{cases} 
1, & (x, y) \in \textit{central region} \\
s, & \textit{otherwise}
\end{cases}
\end{equation}

 Term \textit{central region} corresponds to the low-frequency components in the frequency domain: 
$
x \in \left( \frac{1}{4}W, \frac{3}{4}W \right) \text{ and } y \in \left( \frac{1}{4}H, \frac{3}{4}H \right)
$, while the rest of the frequency domain is scaled by \( s \) to enhance the high-frequency components.

\textbf{For the backbone features}, we apply a two-stage adjustment. In the first stage, we use the structure-aware scaling method proposed by FreeU to enhance the backbone's denoising capability in U-Net. This approach adaptively amplifies features based on their characteristics. Specifically, we compute the channel-wise mean \(\bar{x}\) of the backbone feature \(x\):

\begin{equation}
\bar{x} = \frac{1}{C} \sum_{i=1}^{C} x_{i},
\end{equation}

where \(x_{i}\) is the \(i\)-th channel, and \(C\) is the total number of channels. We then define a scaling map \(\alpha\) as:

\begin{equation}
\alpha = (m - 1) \cdot \frac{\bar{x} - \min(\bar{x})}{\max(\bar{x}) - \min(\bar{x})} + 1,
\end{equation}

where \(m > 1\) is a scalar constant to amplify the backbone. The backbone feature is then strengthened by element-wise multiplication with \(\alpha\):

\begin{equation}
x' = x \odot \alpha.
\end{equation}

In the second step, we aim to reduce the over-smoothing effect caused by enhancing the backbone. Based on previous experiments, which highlighted the role of high-frequency components in the backbone, we suppress these components using a parameter \(b < 1\). This process, similar to the operation on the skip connection, helps improve the texture details of the generated Mel-spectrogram.

Overall, since our method is applied only on the first two decoder blocks, we can achieve optimal results by appropriately setting \(s_1\), \(s_2\), \(b_1\), \(b_2\), and \(m\) during model inference process.

\begin{table*}[h]
\caption{Results Comparison}
\begin{center}
\begin{tabular}{|c|c c c c|c c c c|c|}
\hline
\multirow{2}{*}{\textbf{Model}} & \multicolumn{4}{c|}{\textbf{Original Model}} & \multicolumn{4}{c|}{\textbf{Mel-Refine (Ours)}} & \multirow{2}{*}{\textbf{Setting}} \\
\cline{2-9}
& \textbf{FD} \textcolor{green}{$\downarrow$} & \textbf{FAD} \textcolor{green}{$\downarrow$} & \textbf{KL} \textcolor{green}{$\downarrow$} & \textbf{OVL} \textcolor{red}{$\uparrow$} & \textbf{FD} \textcolor{green}{$\downarrow$} & \textbf{FAD} \textcolor{green}{$\downarrow$} & \textbf{KL} \textcolor{green}{$\downarrow$} & \textbf{OVL} \textcolor{red}{$\uparrow$} & \\
\hline
Tango & 28.36 & 1.73 & 2.15 & 45.5 & \textbf{27.78} & \textbf{1.63} & \textbf{2.12} & \textbf{55.5} & s1=1.2 s2=1.2 b1=0.8 b2=0.1 m=1.4 \\
\hline
MusTango & 25.76 & 1.67 & 1.11 & 41.0 &  \textbf{25.06} & \textbf{1.61} & \textbf{1.04} & \textbf{59.0} & s1=1.4 s2=1.2 b1=0.8 b2=0.6 m=1.1 \\
\hline
Tango2 & 37.48 & 2.48 & 2.31 & 36.0 & \textbf{28.13} & \textbf{1.69} & \textbf{2.11} & \textbf{64.0} & s1=1.4 s2=1.2 b1=0.5 b2=0.1 m=2.5 \\
\hline
\end{tabular}
\label{tab_combined}
\end{center}
\end{table*}

\section{Experiment}
\subsection{Model and Dataset}
To demonstrate the generalizability of our method, we evaluated it using the three latest TTA models across two datasets. Specifically, we tested Tango \cite{tango} and Tango2 \cite{majumder2024tango} on the Audiocaps \cite{kim2019audiocaps,audioset} test set, and MusTango \cite{mustango} on the MusicBench \cite{mustango} test set B. For each experiment, we established a baseline using the model's original inference metrics and then applied our proposed \emph{Mel-Refine} method during inference for comparison. Importantly, the model's denoising steps and guidance parameters remained consistent with those reported in the original papers to ensure a fair comparison.

In preparing the evaluation sets, we randomly selected one text caption per audio clip for Audiocaps—as each clip has five captions—to serve as the text condition. For MusicBench, we followed the same procedure described in MusTango \cite{mustango}. By fixing the random seed during inference, we ensured reproducibility and eliminated variability due to randomness. This approach allowed us to effectively demonstrate the improvements achieved by our method using only a subset of the data, rather than testing the entire dataset.

\subsection{Evaluation Metrics}
\textbf{Objective Metrics:} In this work, we used three commonly used objective metrics: Frechet Distance (FD)\cite{yang2023diffsound}, Frechet Audio Distance (FAD)\cite{FAD}, and Kullback-Leibler Divergence (KL)\cite{audioldm1}\cite{kreuk2023audiogen}. These three metrics have been widely adopted in previous TTA models for evaluating the quality of generated samples. FAD is adapted from Frechet Inception Distance (FID) and measures the distribution-level gap between generated and reference audio samples. KL divergence is an instance-level reference-dependent metric that measures the divergence between the acoustic event posteriors of the ground truth and the generated audio sample. FD is similar to FAD but uses a different feature extraction model.

\textbf{Subjective Metrics:}
We paid ten evaluator to listen to 30 randomly selected audio clips with rich content, comparing and evaluating two sets of audio samples: one generated by the model without using the ``Mel-Refine'' method and the other with it. The evaluation focused on overall audio quality (OVL), and each evaluator voted on both sets of audio samples for each comparison.

\subsection{Model Setting Searching}
Given that we have five parameters to adjust, selecting the optimal parameters requires numerous experiments. We used the following strategy to determine the optimal parameters: first, we determined the backbone enhancement parameter m. This was done by fixing b and s, initially changing m on a large scale to determine the range, and then performing a grid search to find the exact value of m. After determining the value of m, we first verified the relative sizes of b1 and b2, as well as s1 and s2. Generally, $s1 > s2$,$b1 > b2$ yields the best results. Afterward, a grid search was conducted to determine the optimal parameters. The parameters provided in the subsequent results are the optimal ones obtained through experiments.

\subsection{Result and Analysis}


Tables \ref{tab_combined} illustrates the performance improvements achieved by our approach. The table headings reflect the optimal parameter settings identified through experimentation. Overall, all three models exhibited varying degrees of enhancement in objective metrics after applying ``Mel-Refine''. Notably, the Tango2 model demonstrated a nearly 30\% improvement, approaching the performance of Tango1, with FD=28.13 and FAD=1.69. In subjective evaluations, the majority of volunteers reported enhanced audio quality following the application of ``Mel-Refine''.

It is particularly noteworthy that the improvement in the Tango2 model significantly exceeds that of the other two models.This may be due to the fact that Tango2 used DPO\cite{Dpo} fine-tuning, which further aligns the generated audio with the text, but at the cost of reducing the model’s overall generation capability. Our method effectively restored the model’s ability to generate high-quality audio. For the other two models, the smaller improvements in objective metrics might be attributed to their initial optimization being more focused on generating audio that closely resembles real audio, thus leaving limited room for improvement in these metrics. However, in terms of subjective perception, the results indicate that our method indeed enhances the quality of the generated audio.

\begin{figure}[t]
    \centering
    \includegraphics[width=\columnwidth]{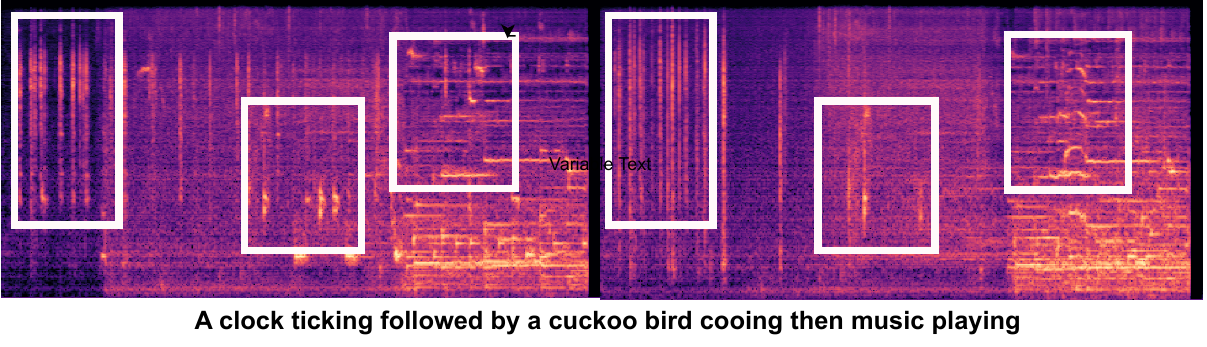} 
    \caption{The left Mel-spectrogram shows the original output from the model, while the right Mel-spectrogram displays the result after applying Mel-Refine. The comparison shows that the clock chime at the beginning is more abrupt in the left Mel-spectrogram compared to the right. In the middle section with the cuckoo bird, the right Mel-spectrogram clearly presents two distinct cooing sounds, and in the final music section, the right Mel-spectrogram also reveals a certain rhythm.}
    \label{fig:final}
\end{figure}

From the generated Mel-spectrogram, our method has significantly refined the details and textures, producing audio that is more realistic and rich in content, as shown in Fig. \ref{fig:final}.

\subsection{Ablation Study}
To evaluate the impact of each key component in our method, we conducted ablation experiments using the Tango2 model. We removed the effects of three specific modules by setting their corresponding parameters to 1.0, while keeping the parameters of all other components unchanged. These modules include skip-connection high-frequency amplify, backbone structure-relating scaling, and backbone high-frequency attenuate. This approach allows us to observe how the performance changes when each module is effectively neutralized. The result is shown in Table \ref{tab3}.
\begin{table}[h]
\caption{Ablation Study Result}
\begin{center}
\resizebox{\columnwidth}{!}{ 
\begin{tabular}{|c|c|c|c|}
\hline
\textbf{Ablate Module} & \textbf{FD} \textcolor{green}{$\downarrow$} & \textbf{FAD} \textcolor{green}{$\downarrow$}& \textbf{KL} \textcolor{green}{$\downarrow$}\\
\hline
Baseline  & 28.13 & 1.69 & 2.11 \\
\hline
Skip-connection high-frequency amplify & 28.66  & 1.71 & 2.14 \\
\hline
Backbone structure-relating scaling & 34.57  & 2.10  & 2.25\\
\hline
Backbone high-frequency attenuate & 28.98 & 2.61 & 2.04 \\
\hline
\end{tabular}
} 
\label{tab3}
\end{center}
\end{table}

The ablation study clearly illustrates the significant contributions of each key component in our method to the overall performance enhancement. The \textbf{backbone structure-relating scaling} module exhibits the most substantial impact, particularly on the FD metric, emphasizing its critical role in preserving structural consistency and global features. The \textbf{backbone high-frequency attenuate} module notably improves the FAD metric, highlighting its importance in refining audio quality and detail. Although the \textbf{skip-connection high-frequency amplify} module has a comparatively smaller effect, it still contributes to the nuanced enhancement of high-frequency information, aiding in the capture and retention of subtle details. These findings underscore the necessity of integrating all three modules to achieve optimal results, as they collaboratively balance the amplification and attenuation of high-frequency components while maintaining the overall quality and coherence of the generated outputs.

\section{Conclusion}
In conclusion, we presents a novel approach, ``Mel-Refine'', which enhances the quality of Mel-spectrograms generated by diffusion-based TTA models. By analyzing the roles of high-frequency and low-frequency components in the U-Net architecture, we demonstrate that high-frequency components in skip connections and the backbone contribute to the texture and detail of generated audio, while low-frequency components are critical for the denoising process. Our proposed method requires no additional training or fine-tuning, making it highly generalizable and compatible with existing diffusion-based architectures. Experimental results show that ``Mel-Refine'' effectively improves the clarity and coherence of generated audio across various models and datasets, advancing the field of TTA audio generation.

\newpage
\balance
\bibliographystyle{IEEEtran}
\bibliography{ref}

\end{document}